\documentclass[12pt]{article}
\input epsf
\newcommand{\ba}{\begin{array}}
\newcommand{\ea}{\end{array}}

\newcommand{\be}{\begin{equation}}
\newcommand{\ee}{\end{equation}}
\newcommand{\bea}{\begin{eqnarray}}
\newcommand{\eea}{\end{eqnarray}}
\newcommand{\beal}{\setcounter{letter}{1} \begin{eqnarray}}
\newcommand{\eeal}{\addtocounter{equation}{1} \end{eqnarray}}

\newcommand{\req}[1]{Eq.(\ref{#1})}

\newcommand{\larrow}{\,\,\,\,\hbox to 30pt{\rightarrowfill}
\,\,\,\,}
\newcommand{\slarrow}{\,\,\,\hbox to 20pt{\rightarrowfill}
\,\,\,}

\newcommand{\half}{{1\over2}}

\newcommand{\bm}{\bibitem}

\usepackage[dvips]{graphicx}
\newcommand{\vtab}{\vspace{.5 cm}}
\newcommand{\vhtab}{\vspace{.25 cm}}
\usepackage[dvips]{graphicx}

\newcommand{\tab}{\hspace{.5 cm}}
\newcommand{\htab}{\hspace{.2 cm}}

\newcommand{\beqn}{\begin{equation}}
\newcommand{\eeqn}{\end{equation}}

\begin{document}

\begin{titlepage}
\renewcommand{\thefootnote}{\fnsymbol{footnote}}
\renewcommand{\baselinestretch}{1.3}
\medskip
\hfill  UNB Technical Report 06-02\\[20pt]

\begin{center}
{\large {\bf Quantum Structure of Space Near a Black Hole Horizon }}
\\ \medskip  {}
\medskip

\renewcommand{\baselinestretch}{1}
{\bf
J. Gegenberg $\dagger$,
G. Kunstatter $\sharp$,
R.D. Small $\flat$,
\\}
\vspace*{0.50cm}
{\sl
$\dagger$ Dept. of Mathematics and Statistics and Department of Physics\\
University of New Brunswick\\
Fredericton, New Brunswick, Canada  E3B 5A3\\ [5pt]
}
{\sl
$\sharp$ Dept. of Physics and Winnipeg Institute of
Theoretical Physics\\
University of Winnipeg\\
Winnipeg, Manitoba, Canada R3B 2E9\\[5pt]
}
{\sl
$\flat$ Dept. of Mathematics and Statistics\\
University of New Brunswick\\
Fredericton, New Brunswick, Canada  E3B 5A3\\ [5pt]
}
\end{center}

\renewcommand{\baselinestretch}{1}

\begin{center}
{\bf Abstract}
\end{center}

We describe a midi-superspace quantization scheme
for generic single
horizon black holes in which only the spatial diffeomorphisms are fixed. The remaining Hamiltonian constraint yields an infinite set of decoupled eigenvalue equations: one at each spatial point. The corresponding operator at each point is
the product of the outgoing and ingoing null convergences, and describes the scale invariant quantum mechanics of a particle moving in an attractive $1/X^2$ potential. The variable $X$ that is analoguous to particle position is the square root of the conformal mode of the metric. 
We quantize the theory via 
Bohr quantization, which by construction turns the Hamiltonian constraint 
eigenvalue equation into a finite difference equation. The resulting spectrum gives rise to a discrete spatial 
topology  exterior to the 
horizon. The spectrum approaches the continuum in the asymptotic region. 

{\small}
\vfill
\hfill  
\today
\end{titlepage}
\section{Introduction}
It is commonly believed that the puzzling thermodynamic properties of black holes have their origin in the quantum behaviour of the gravitational field.
The two leading candidates for a consistent quantum theory of gravity each have their own picture of the structure that underlies spacetime geometry. In the case of string theory, the fundamental microscopic objects are supersymmetric strings and branes, whereas in loop quantum gravity  the fundamental 
structures are spin networks that generate a quantum phase 
space whose coordinates are holonomies on the spin networks and densitized 
spatial triads. Although both theories have made progress within their respective frameworks\footnote{See for example \cite{big bang}, in which it is shown that loop quantum cosmology is capable of resolving the big bang singularity}, it is safe to say that a great deal remains to be learned, particularly about the quantum gravitational origins of the thermodynamic properties of black holes.

One striking aspect of black hole thermodynamics is the apparent universality: the key features seem not to depend on the specific action from which they are derived. All that is required is the presence of an event horizon in a diffeomorphism invariant theory. It therefore makes sense to try to understand the quantum behaviour of black holes, and the resulting thermodynamics not from the point of view of a specific microscopic underlying theory, but instead in terms of the things that they have in common (at the semi-classical level at least), namely their geometrodynamic properties. 

Recently Husain and Winkler have initiated a program\cite{hw1} designed to 
address directly the quantum geometrodynamics of black holes.  Their viewpoint is that returning to 
geometrodynamical variables  opens up 
new possibilies for resolving the Hamiltonian constraint.  
Of course, this cannot be accomplished by a return to the original 
formulation, wherein the 
quantum phase space was parametrized by the spatial metric and its 
conjugate momentum.  Instead, adapting a quantization method used recently in the context of loop quantum gravity(see \cite{shadow}), the latter is replaced by its exponentiated 
version.  This necessitates the construction of a quantum phase space with 
a {\it discrete} topology, as is the case with `conventional' loop quantum gravity.  

At the present time, this program has only been fully implemented for `reduced'
systems, that is, for the case of dynamics of homogeneous and isotropic 
universes, or where there is spherical symmetry.  Since a good 
part of gravitational physics deals with these two cases, there have already 
been interesting results regarding singularity resolution and in understanding 
the quantum definition of horizons \cite{hw1}.  The first steps have also been taken towards implementation of the program in a partially reduced theory with matter \cite{hw3}.

In this paper, we will, first of all, broaden the applicability of this program by applying it to generic single horizon black holes using the formalism of two dimensional dilaton gravity models.  This extends 
the program to spherically symmetric gravity in arbitrary dimensions, as well as to models of interest in  string theory, such as the CGHS model of two 
dimensional gravity\cite{cghs}.  

Second of all, we will use a method of gauge fixing distinct from \cite{hw3} that turns out to have interesting properties as well as potential implications beyond the Bohr quantization program and the vacuum case considered here. In particular, we look at a midi-superspace model obtained by partially fixing the spatial diffeomorphisms, so that in effect, the spatial coordinate is a 
specified function of the dilaton field. Our partial gauge fixing is very natural from the geometrodynamics viewpoint and allows one to examine the evolution of spatial slices that extend from the singularity across the horizon and off to infinity.  Remarkably, the resulting Hamiltonian operator reduces to the product of the ingoing and outgoing null convergences at each spatial point. Moreover, the Hamiltonian at each point corresponds to that of a particle moving in a $1/X^2$ potential. The $1/X^2$ potential is scale invariant and its quantization has been studied extensively in the past\cite{ordonez,camblong1}, both for its interesting mathematical properties and for its physical relevance. Most significantly, it arises in the near horizon dynamics of black holes where the conformal symmetry has been used to account for the entropy associated with the horizon\cite{carlip, solodukhin}. (See also \cite{camblong2} for an analysis of the connection between this potential and black hole thermodynamics.)

We recognize the relative triviality of vacuum dilaton gravity- there are 
no propogating degrees of freedom. Nevertheless, we will show that the 
implementation of the Bohr quantization program in this context allows us to solve the Hamiltonian 
constraint, and sets the stage for our consideration of quantum evolution of 
systems containing gravity and propagating  matter. Moreover, if indeed 
diffeomorphisms near the boundary do play a role in black hole entropy, as suggested by Carlip\cite{carlip} and others, it is of great interest to able to quantize these modes as well as the physical ones. Finally, and perhaps most significantly, our quantization scheme has interesting consequences even for black hole vacuum spacetime: it induces a discrete spatial topology exterior to the horizon that approaches the continuum at large distances.

\section{Classical Theory}

We start from the action 
\be
S[g,\phi]=\frac{1}{2G}\int dx dt \sqrt{-g}\left(\phi R(g)+{V(\phi)\over l^2}\right).\label{dgaction}
\ee
In the above $G$ is the two dimensional Newton's constant, which has units of $1/\hbar$, while $l$ is an arbitrary length scale that we take to be the Planck scale for the theory. This model has been extensively studied in the past using a variety of formalisms (see \cite{2dreview} for a recent review). 
This theory is exactly solvable both classically and quantum mechanically. For details of the analysis in terms of geometrodynamic variables, see \cite{exact}. 

This action describes, for different choices of dilaton potential, a very large class of theories. Spherically symmetric Einstein gravity in four spacetime dimensions corresponds to a dilaton potential $V \propto 1/\sqrt{\phi}$ and 
on shell $\phi = r^2$ is the area of symmetric spheres of radius $r$. For the more general case of $d$-dimensional Einstein gravity with $d-2$ dimensional spherical symmetry, $V\propto \phi^{-1/(d-2)}$, $\phi\propto r^n$ is the area of a unit n-sphere and $G = 8\pi G^{(d)}/ (l^{(d-2)}\Omega^{(d)})$, where $\Omega^{(d)}$ is the volume of the unit $n$-sphere. One can without loss of generality take $l=(\hbar G^{(d)})^{1/(d-2)}=l_{pl}$ to be the higher dimensional Planck length.

The vacuum theory admits a one parameter family of solutions with at least one Killing vector. This solution can be represented in Schwarzschild like coordinates $x=l \phi$ as:
\be
ds^2= -(j(\phi) -2GMl)dt^2+(j(\phi)-2GMl)^{-1}dx^2,
\ee
where $j(\phi) = \int^{\phi} d\phi V(\phi)$ and $M$ is the ADM mass. For the present purposes it is
convenient to assume that $j(\phi)$ is a monotonic function of $\phi$, in which case each vacuum solution contains a horizon located at $\phi_h = j^{-1}(2GMl)$

The canonical analysis follows that in \cite{exact}. The metric is written in modified ADM form:
\be
ds^2=e^{2\rho}\left(-\sigma^2 dt^2+\left(dx+N dt\right)^2\right).
\label{metric}
\ee
We are assuming that the 2D spacetime manifold $M_2=R\times\Sigma$, where $\Sigma$ is a spacelike 
slice.   As usual the lapse and shift functions $\sigma$ and $N$, respectively,
 are Lagrange multipliers, while the momenta conjugate to the fields $\phi$ 
and $\rho$ defined above are:
\bea
G\Pi_\phi &=& {1\over \sigma} (N\rho'+N' - \dot{\rho}),\nonumber \\
G\Pi_\rho &=& {1\over \sigma}(N\phi'-\dot{\phi}),
\eea
and the hamiltonian is
\be
H=\int \left(\sigma\cal G +N\cal F\right)+\hbox{surface terms}.
\ee
Thus the Lagange multipliers $\sigma,N$ enforce the secondary constraints (respectively the Hamiltonian and diffeomorphism constraints):
\bea
G\cal G:&=&-G\Pi_\rho G\Pi_\phi+\phi''-\phi'\rho'-\half e^{2\rho}{V(\phi)\over l^2},\\
\cal F:&=&\rho'\Pi_\rho-\Pi_\rho'+\phi'\Pi_\phi.
\eea
The momenta canonically conjugate to $\sigma,N$ are primary constraints.

An improved Hamiltonian constraint is 
\be
\tilde{\cal G}:=-le^{-2\rho}\left(\phi'{\cal G}+G\Pi_\rho{\cal F}\right).
\ee
It turns out that $\tilde{\cal G}=\cal H'$  is a total spatial derivative , with
\be
{\cal H}:={l\over 2G}\left[ e^{-2\rho}\left({G^2\Pi^2_\rho}-(\phi')^2\right)+
{j(\phi)\over l^2}\right], 
\ee
on the constraint surface. It follows that the 
Hamiltonian constraint is equivalent to
\be
{\cal H}=M=constant.
\label{cal H1}
\ee
where $M$ is the energy, or black hole mass.

We now perform a canonical transformation:
\bea
X&=& e^\rho,\\
P&=&{e^{-\rho}\Pi_\rho},
\eea
so that 
\bea
\cal F&=&-XP'+\phi'\Pi_\phi,\\
\cal H&=&{l\over 2G}\left[G^2P^2-\frac{(\phi')^2}{X^2}+ {j(\phi)\over l^2}\right].
\label{improved ham}
\eea
\section{The Classical Null Expansion Observable}
We now verify an important property of the improved Hamiltonian constraint
operator (\ref{improved ham}), namely its relationship to the expansion
of null rays. Given that we are working in two spacetime dimensions, the 
definition of expansion requires a bit of care. Guidance is provided by the fact that one can consider 2D dilaton gravity as dimensionally reduced Einstein gravity in $D=n+2$ spacetime dimensions. In this case the dilaton has the interpretation as the invariant area of the $n-$sphere spanned by the generators of spatial rotations, i.e. the area of a sphere of fixed radius $r$. Even in the case where the 2D theory is considered fundamental, the dilaton provides the only scalar field whose rate of change along null vectors has an invariant meaning. With this motivation we define the expansion $\Theta^\pm$ as the fractional change in the value of the dilaton along ingoing and outgoing null vectors, respectively:
\be
\Theta^{(\pm)} = {\nabla^\mu \phi\over\phi}l^{(\pm)}_\mu,
\ee
where the vectors
\be
l^{(\pm)}_\mu = n_\mu \pm s_\mu,
\ee 
are ingoing and outgoing null vectors defined in terms of the future pointing
normal $n_\mu = e^\rho[\sigma,0]$ of each spacelike slice defined by the 
metric (\ref{metric}), and the embedded inward pointing normal
$s_\mu=e^\rho[-N,-1]$.\\

A straightforward calculation then gives:
\bea
\Theta^{(\pm)}&=& {e^{-\rho}\over \phi}\left({\dot{\phi}\over\sigma}+
   \phi'\left({N\over\sigma} \pm {1} \right)\right)\nonumber\\
  &=& {e^{-\rho}\over\phi}(\Pi_\rho \pm \phi')\nonumber\\
  &=&{1\over\phi}\left(P\pm {\phi'\over X}\right).
\eea

As in Husain and Winkler\cite{hw5}, the expansions provide phase
space functions that allow one to test for the presence of horizons.
We now see that the Hamiltonian operator can be written in terms of the
expansion observables:
\be
{\cal H} \propto \left[\Theta^+ \Theta^- + j(\phi)\right].
\ee

\section{Gauge Fixing}
It is well-known that 2D dilaton gravity has no propogating degrees of 
freedom.  This follows immediately from the fact that the Hamiltonian is a 
linear combination of the secondary constraints $\cal F,\cal G$.  Thus 
we in can principle reduce our system to a quantum mechanical system by gauge fixing. Here, we fix only the freedom associated with spatial diffeomorphisms by choosing
\be
\chi(x):=\phi'- f(\phi)/l= 0.\label{gauge}
\ee
which implicitly specifies the spatial coordinate in terms of the invariant
dilaton. Recall that $\phi$ is the invariant area of a symmetric
n-sphere in spherically symmetric gravity in d=n+2 dimensions. At this 
stage any monotonic function $f(\phi)$ would do but we would like to describe the quantum mechanics of a black hole spacetime inside, outside and in the neighbourhood of the horizon. We therefore make the choice
$f(\phi)= j(\phi)$, which is consistent with the 2D dilaton gravity anologue of Painleve-Gullstrand coordinates, in which the generic metric takes the
form:
\be
ds^2 = j(\phi)\left( -dt^2 + (dx + \sqrt{2GMl\over j}dt )^2\right),
\ee
where, from (\ref{gauge}), $dx =l d\phi/j(\phi)$. 

When we smear the diffeomorphism constraint with a test function that vanishes on the boundaries:
\be
\chi(\lambda)\equiv \int_\Sigma dx \chi(x)\lambda(x),
\ee
 and compute its Poisson bracket with the constraint, we find
\be
\left\{\chi(\lambda),{\cal F}(y)\right\}= \int dx \tau j(\phi)\left(\lambda' + {dj(\phi)\over d\phi}\lambda\right).
\ee
Given the boundary conditions on $\lambda(x)$ this bracket is invertible.
Hence, the gauge choice above is acceptable, in the Dirac sense.

The consistency condition that preserves the gauge fixing condition in time is:
\be
\sigma G\Pi_\rho = N \phi',
\ee
and one can now eliminate $\Pi_\phi$ from the dynamical equations by solving the diffeomorphism 
constraint.
The partially reduced Hamiltonian is:
\be
H = -\int_\Sigma dx\tilde{\sigma}{\cal H}' + \left.\tilde{\sigma}{\cal H}
\right|_{boundary}.
\label{final hamiltonian}
\ee
It is easy to show that the Dirac bracket of $X$ with $P$ reduces to the Poisson bracket of 
these variables so that the equations of motion for the remaining  fields can be derived directly from (\ref{final hamiltonian}) using the ordinary Poisson brackets. In this gauge, the theory is determined entirely by the
function ${\cal H}$, which on the constraint surface is a constant of
motion. The Hamiltonian constraint (\ref{cal H1}) now takes the simple form:
\be
O(X,P)\equiv (G l P)^2-\frac{j(\phi)^2}{X^2}-\left(2GMl-j(\phi)\right)=0.
\label{ham1}
\ee

Since we have fixed the spatial diffeomorphisms, the value of $\phi$ at each spatial point
can be considered a c-number. Thus only $P$ 
and $X$ are to be quantized. In this gauge, there is no coupling between 
different spatial points so that, remarkably, the midi-superspace model has decoupled into an infinite set of quantum mechanical models. 
Moreover, for each value of $\phi$, the Hamiltonian in (\ref{ham1}) is that of a particle moving in a $1/X^2$ binding potential with  total ``energy'' 
\be
\epsilon \equiv 2GMl-j(\phi),
\ee
determined 
by the black hole mass and the invariant location along the spatial slice as determined by the value of $\phi$. The 
total ``energy'' is negative in the asymptotic region outside the horizon, 
so that one is looking for bound states in this region, whereas the energy 
is positive in the interior where one is looking for ``scattering states''. 
At the classical horizon location $ j(\phi)=2 G M l$, the eigenvalue is zero. 

The $1/X^2$ potential arises in a variety of contexts, including the quantum Hall effect\cite{caval} and a field theory formulation of nucleon interactions \cite{weinberg}. More importantly for the present context it arises in the near horizon dynamics of black holes where the conformal symmetry has been used to account for the entropy associated with the horizon\cite{carlip, solodukhin}. 

An important property of the inverse square potential Hamiltonian (\ref{ham1}) is its scale invariance. As a result, it possesses three symmetry generators that span an algebra isomorphic to the conformal group $SO(1,2)$. The quantum mechanics of this potential has been the subject of much work, in part because the conformal symmetry is necessarily broken at the quantum level. An elegant way of understanding the anomaly was given by Esteve\cite{esteve}: the other generators of the conformal group do not preserve the domain of self adjointness of the Hamiltonian, which resulting non-self-adjointness leads directly to the anomalous term in the quantum algebra.  One can also reproduce this anomaly by introducing a regulator (either a cut-off  
dimensional regularization\cite{camblong1}), and verifying that the algebra does not close even after the regulator is removed.

In the present paper, we will use a regularization that seems appropriate in the quantum gravity 
context, namely Bohr quantization \cite{shadow,hw1}.

\section{Bohr Quantization}
We now quantize the partially gauge-fixed theory in the discrete Bohr quantization framework \cite{shadow,hw1}. We first smear the observable $X(x)$ with a test function $f(x)$:
\be
X_f\equiv\int_\Sigma dx f(x) X(x).
\ee
Secondly we define a basis $|a_1,a_2, ...., a_N>$ of orthonormal states, such that:
\be
<a'_1,a'_2...a'_N|a_1,a_2,...,a_N> = \delta_{a'_1,a_1}\delta_{a'_2,a'_2}....,
\ee
on which we define the action of the quantum smeared operator $\hat{X}_f$:
\be
\hat{X}_f |a_1,a_2, ...., a_N> = L\sum_i (f(x_i)a_i)|a_1,a_2, ...., a_N>.
\label{Xhat}
\ee
where $L$ is a constant of dimension length which will be determined below. The discretization $\{x_i\}$ is at this stage completely arbitrary. The $a_i$ are the eigenvalues of $\hat{X}$ at the point $x_i$. This spectrum is, by construction, discrete, and will be determined once we choose a representation for its conjugate
momentum, $\hat{P}$. 

Instead of quantizing $\hat{P}$ directly we first of all note that it is the dimensionless 
variable $GlP$ that appears in the Hamiltonian constraint. We therefore define:
\be
U_\gamma(x) \equiv \exp(i\gamma G l P(x)),
\ee
where the constant $\gamma$ is dimensionless. The Poisson bracket is:
\be
\{X_f,U_\gamma(x)\} = i\gamma G l f(x) U_\gamma(x).
\label{poisson bracket}
\ee
We define the quantum version of $U_\gamma(x)$ by its action on the basis states: 
\be
\hat{U}_\gamma(x_i)|a_1,a_2, ...., a_N>=|a_1,...,a_i-\gamma, ...., a_N>.
\label{Uhat}
\ee
The definitions (\ref{Xhat}) and (\ref{Uhat}) imply the quantum commutator:
\be
[\hat{X}_f, \hat{U}_\gamma(x_i)]=-\gamma L \hat{U}_\gamma(x_i).
\label{commutator}
\ee 
A comparison of (\ref{poisson bracket}) and (\ref{commutator}) determines the arbitrary constant $L$ to be proportional to the Planck scale:
\be
L=\hbar G l.
\ee

We are now in a position to determine the spectrum of $\hat{X}$. In particular, given the action of 
$\hat{U}_\gamma(x_i)$, one has that:
\be
|a_1,...,a_i,...,a_N> = (\hat{U}_{\pm\gamma}(x_i))^m|a_1,...,a^{(0)}_i,...,a_N>=|a_1,...,a^{(0)}_i\pm m\gamma,...,a_N>.
\ee
Given the physical interpretation of $X$ in terms of the square root of the conformal mode of the metric, it is reasonable to assume that the spectrum is symmetric about 0, and take $a_i^{(0)}=0$. Henceforth we assume that $a_i = m\gamma$ where $m$ is any integer. 

We implement the Hamiltonian constraint using the following smeared operator:
\bea
\hat{O}_f X^2|a_1,...., a_N>\nonumber&=&
  {1\over 2}\int dx f(x)\left(l^2 G^2\hat{P}^2\hat{X}^2-{j(\phi)^2}
-{\epsilon(\phi)}\hat{X}^2\right)|a_1,...., a_N>\nonumber\\
&\equiv&{L\over 2}\sum_i f(x_i)\left(a_i^2G^2l^2\hat{P}^2 -{j_i} -
{\epsilon_i}a_i^2\right)|a_1,...., a_N>.
\label{Hhat}
\eea
where for ease of notation we henceforth denote $j_i:=j(\phi(x_i))$ and 
similarly for $\epsilon_i$.

We now need to define $\hat{P}^2(x_i)$. Following Ashtekar {\it et. al.} \cite{shadow}:
\bea
l^2\hat{P}^2(x_i)|a_1,...,a_N>&\equiv& \lim_{\gamma\to 0}{1\over \gamma^2 }\left(
2-U_\gamma(x_i)-U_{-\gamma}(x_i)\right)|a_1,...,a_N>\nonumber\\
&=&\lim_{\gamma\to0} {1\over \gamma^2}\left(
  2 |a_1,...,a_N> - |a_1,...,a_i-\gamma,..,a_N> \right.\nonumber\\
&-& \left.|a_1,...,a_i+\gamma,a_N> \right).
\label{P2hat}
\eea

We are now ready to solve the Hamiltonian constraint
\be
\hat{O}_f X^2|\psi> = 0,
\label{quantum constraint}
\ee
where $|\psi>$ an arbitrary state in the Hilbert space:
\be
|\psi> = \sum_{a_1,...,a_n}\psi(a_1,...a_N)|a_1,...,a_N>.
\label{arb state}
\ee
Since there is no coupling between spatial points in the Hamiltonian constraint, it is reasonable to separate variables:
\be
\psi(a_1,...a_N) = \psi_1(a_1)\psi_2(a_2)...\psi_N(a_N).
\label{ansatz}
\ee
By substituting (\ref{arb state}) and (\ref{ansatz}) into (\ref{quantum constraint}) and using the definitions (\ref{Hhat}) and (\ref{P2hat}) one finds that the coefficients $\psi_i(a_i)$ must each satisfy:
\be
{a_i^2\over \gamma^2 }\left( \psi_i(a_i+\gamma)+ \psi_i(a_i-\gamma)\right)
=\left({2\over \gamma^2 } a_i^2 - \epsilon_i a_i^2
    - j_i^2\right)\psi_i(a_i).
\label{difference equation 1}
\ee

A normalizable state $|\psi>$ satisfies
\be
<\psi>^2:=\sum_m |\psi^m|^2<\infty.
\ee
We henceforth suppress the index $i$ labelling the spatial point.
As in  \cite{shadow} we use the fact that $a_i=m\gamma$ to write $\psi(a_i)=\psi^m$ so that
 $\psi(a_i\pm\gamma)=
\psi^{m\pm1}$. Hence (\ref{difference equation 1}) can be written as:
\be
(m\gamma)^2\left[\psi^{m+1}+\psi^{m-1}
-\left(2-\epsilon\gamma^2\right)\psi^m\right]+\gamma^2 j^2
\psi^m=0.
\label{difference equation 2}
\ee

To best understand the issue of convergence, we follow the lead of 
Ashtekar et. al. \cite{shadow} and Fourier transform in $m$ to a continuous 
variable $k$, so that the difference equation becomes a second order 
differential equation.  

Multiplying (\ref{difference equation 2}) by $\exp(ikm\gamma)$ and 
summing over $m$, and using
\be
\psi(k):=\sum_m\psi^m e^{imk\gamma}, 
\ee
so that
\be
\sum_m(m\gamma)^2\psi^{m-1}e^{imk\gamma}=-\partial^2_k\left(e^{-ik\gamma}
\psi(k)\right),
\ee
and similarly for $\psi^{m+1}$, we get
\be
\partial^2_k\left[\left(1-\cos{(k\gamma)}-\epsilon\frac{\gamma^2}{2}
\right)\psi(k)\right]+\frac{\gamma^2 j^2}{2}\psi(k)=0.\label{bohrde}
\ee

Now choose a new variable $z:=\gamma k/2$.  Then the above can be written as
\be
(1+A \cos{2z})\psi''(z)+B\sin{2 z}\psi'(z)+(C+D\cos{2z})\psi(z)=0,\label{ince}
\ee
where the `constants'- that is, parameters independent of $z$, are defined by 
\bea
A&:=&-(1-\epsilon(\phi)\gamma^2/2)^{-1},\\
B=D&:=&4(1-\epsilon(\phi)\gamma^2/2)^{-1},\\
C&:=&2(j(\phi))^2(1-\epsilon(\phi)\gamma^2/2)^{-1}.
\eea
The differential equation \req{ince} is recognized as {\it Ince's differential equation} \cite{m&w} .  
\footnote{In \cite{shadow} the
non-relativistic simple harmonic oscillator is quantized in the above manner,
and the differential equation corresponding to \req{bohrde} is the Mathieu
equation.}

It is proved in \cite{m&w} that $|A|<1$ is a sufficient condition that there are an infinite number of solutions of 
Ince's equation which are periodic with period $\pi$ or $2\pi$.  By Parseval's Theorem, it follows 
that these periodic solutions of \req{ince} generate normalizable states satisfying the 
Hamiltonian constraint \req{quantum constraint}. The condition $|A|<1$ is satisfied for $j>2GMl$, i.e. exterior to the horizon and when 
\be
j<2GMl - 4/\gamma^2.
\ee
This implies that we are guaranteed the existence of normalizable solutions everywhere exterior to the horizon. However, the width of the region inside the horizon where solutions are not guaranteed is, in terms of the area $\phi_H=j^{-1}(2GMl)$ of the horizon,:
\be
{\Delta \phi\over \phi_H}\sim {4\over \gamma^2}{1\over 2GMl}.
\ee
 This is narrow for macroscopic black holes with mass:
\be
2GMl>{4\over \gamma^2},
\ee
but is of order one for microscopic black holes. In the classical 
limit, $\gamma\to 0$, the entire interior of the black hole is potentially excluded. 

In the next section we will discuss in some detail the spectrum exterior to the horizon of the 
solutions of the difference equation.  First we will present a  
heuristic argument that one expects a classical spacetime geometry to 
emerge in the region far from the event horizon.

Consider the factor 
$a:=1-\epsilon(\phi)\gamma^2/2$ which occurs in the coefficients of \req{ince}. We wish to consider the
region for which $a>>1$, which in turn implies that:
\be
(j-2GMl){\gamma^2\over 2} >>1,
\ee
In this limit the solutions should approximately obey the 
differential equation
\be
\psi''+\frac{4}{a}\sin{(2z)}\psi'+\lambda^2\psi=0,
\ee
where $\lambda:=\sqrt{2/a}j(\phi)$ which is large for 
large $j(\phi)$, so that one can drop the term in $\psi'$. In this case, the 
solutions are of the form
\be
\psi(z)=\psi_0 \sin(\lambda z).
\ee
and the spectrum is
\be
\lambda=2N,
\ee
for some integer $N$. Using the definition of $a$ this gives the following 
equation for $j$ in terms of $N$:
\be
j^2=2N^2\left(1-(2GMl-j){\gamma^2\over2}\right).\label{asyj}
\ee
Under the assumptions made above, the term proportional to $j$ dominates
the right hand side, so that the approximate spectrum far from the black hole is:
\be
j = N^2 \gamma^2.
\label{asymptotic spectrum}
\ee
Using (\ref{asymptotic spectrum}) and the fact that $j>>2GMl$, we find:
\be
{\Delta j\over j} = {2\over N}<<2{\gamma\over \sqrt{2GMl}},
\ee
which is small for macroscopic black holes as expected.

\section{Details of the Spectrum}
We now  establish an asymptotic expansion for each of the 
infinitely many allowed 
values of the spatial coordinate as parametrized by the quantity $j(\phi)$.  
We consider the difference equations \req{difference equation 2}(or, equivalently, the recursion 
relations for periodic solutions of Ince's equation).  These equations are  
rewritten as
\bea \psi^0&=&0,\\  \psi^{n+1}+\psi^{n-1} &=& (2a-\frac{b}{n^2})\psi^n,\, n=1,2,\dots 
\label{consist} \eea
posed as a second order difference equation and one initial condition. In the 
above, 
\bea
a&:=&1+(j(\phi)-2GM\ell)\gamma^2/2;\nonumber\\
b&:=&j^2(\phi). 
\label{ab}
\eea
We will 
obtain an infinite number of self-consistency conditions, fixing the spectrum.

Since the above is homogeneous and linear in the $\psi^n$, 
we may normalize by taking $\psi^1=1$ and then obtain 
a general solution by multiplying \req{consist} by an arbitrary constant.
With these two initial conditions the $\psi^n$ are given for all nonnegative
integers $n$. We require further that the $\psi^n$ converge for large $n$
in order to assert that an infinite set of equations is satisfied. 
The convergence condition is the constraint on $a$ and $b$ that we are seeking.
When we have this condition the $\psi^n$ will be dictated by conditions near
$n=\infty$ and the difference equation then gives all the $\psi^n$. When $\psi^2$
becomes known, then use of the initial conditions and the difference equation
for $n=1$ gives the constraint \htab $\psi^2=2a-b$. \htab The order of the 
difference equation is lowered by noting that it involves only the expression
$\psi^{n+1}/\psi^n =k_n$ and making this substitution leads to the first order
difference equation and initial condition
\bea k_n&=&2a-b/n^2-\frac1{k_{n-1}} , \htab n=2,\dots, \htab\nonumber\\ 
  k_1&=&2a-b. 
\label{consk} 
\eea 
The constraint on $a$ and $b$ is now given by \htab $f(a,b)=0$ where
\beqn f(a,b)=2a-b-k_1(a,b). \label{fab} \eeqn

 From \req{consk} we have 
\beqn k_{n-1}=\frac1{2a-b/n^2-k_n}. \label{knmo} \eeqn
so that the constraint (\ref{fab}) can be written as a continued fraction equation:
\beqn
0 = V_1-{1\over V_2-{1\over V_3-{1\over V_4 ...}}} ,
\eeqn
where:
\beqn
V_n:= 2a - {b\over n^2}.
\eeqn

Let us now  consider the case that $a$ and $b$ are large, both
because this condition leads quickly to the algorithm for finding the 
relation between $a$ and $b$ and because the numerical expression of
this relation suggests what asymptotic expansion to look for. If $n$
is very large then the term $b/n^2$ can be neglected and we find that 
\req{knmo} is, in this limit of large $n$, a very quickly convergent
iteration. If we are to solve $k=g(k)$ by iteration, then a fixed point
is convergent if $g'(k)$ is less than one in size and for the above 
iteration we have rapid convergence. The algorithm for evaluation of 
$f(a,b)$ is to find $k_{n-1}$ for large $n$, given $a$ and $b$ and 
repeatedly substitute until $k_1$ is found. Since $2a-b/n^2$ is 
large and $k_n$ is small we start the iteration with the guess that
$k_n=0$. At the same time, since one generates the sequence of values
of $k_n$, one can obtain a sequence of derivatives
\[ k_{n-1,a} =\frac{k_{n,a}-2}{(2a-b/n^2-k_n)^2}, \htab 
k_{n-1,b} =\frac{k_{n,b}+1/n^2}{(2a-b/n^2-k_n)^2}, \]
again starting the iteration with the derivatives at $n$ being zero.
Newton's method can then be performed on the equation $f(a,b)=0$ and 
the result is plotted in Figure 1. This diagram, it may be noted, has
curves that for large $a$ and $b$ appear to straighten out and their
slopes happen to be $2m^2$ for integers $m=1,2,\dots$.

We now seek an asymptotic expansion for the curves of Figure 1 for 
large $a$ and $b$. We notice that values for $k_n$ propagate 
stably from high to low values of $n$ by substitution into \req{knmo} 
since all $k_n$ are small and the value of $k_{n-1}$ is dominated by
the $a$ and $b$ terms of the denominator.  On the other hand, we note 
from \req{consk} that unless the large $a$ and $b$ terms happen to
cause the difference $2a-b/n^2$ to be small for some integer $n$, 
first, $k_1$ is large due to the initial condition and then $k_n$ is
large for increasing integers $n$. Thus, we obtain a contradiction unless 
$b$ is close to $2am^2$ for some integer $m$. The algorithm for 
solving for the $k_n$ is to find the small $k_n$ for $n \geq m$ and
the large $k_n$ for $n=1, 2, \dots m$. The coefficients for $k_m$ found
both ways must be consistent and their equality gives the conditions
that determine the $b_i$.  

The following substitution happens to give the correct form for the 
solution to the difference equation in \req{consk}.  Set
\beqn k_n=\sum_{i=0}^{\infty}c_i(n)a^{1-2i}, \htab b-2an^2
=\sum_{i=0}^{\infty}b_ia^{1-2i}, \label{subst} \eeqn
where $b_0=2(m^2-n^2)$ and the rest of the $b_i$ are independent
of $n$ and unknown. First we determine the form
of the $k_n$ that propagate in from large values of $n$. Since
$k_n$ is small we know beforehand that $c_0(n)=0$ for $n\geq m$.
Placing the form \req{subst} into \req{knmo}, we have
\[ \sum_{r=0}^{\infty}\sum_{i=0}^{r} \left(c_i(n)c_{r-i}(n-1)
+c_{r-i}(n-1)b_i/n^2 \right) a^{2-2r} +\delta_{1,r}=0.\]
Equating coefficients of each power of $a$ we obtain
\[ c_1(n-1)=\frac1{2(1-m^2/n^2)}, \htab n=m+1,m+2,\dots.\]
\beqn c_r(n-1)=\frac{c_i(n)c_{r-i}(n-1)+c_{r-i}(n-1)b_i/n^2}{2(1-m^2/n^2)},
\htab r=2,3,\dots, \htab n=m+1,m+2,\dots.\label{cs1} \eeqn
We now propagate solutions for $k_n$ beginning at $n=1$.
Substituting \req{subst} into \req{consk} we obtain the result
\beqn c_r(n)=-\frac{\sum_{i=0}^rc_{r-i(n-1)b_i/n^2}-\delta_{r,1}
+\sum_{i=0}^{r-1}c_i(n)c_{r-i}(n-1)}{c_0(n-1)}, \htab r=0,1,2, 
\dots \label{cs2} \eeqn
which may be used for $n=2,3,\dots, m$ to give a second expression
for the coefficients of $k_n$ in addition to that of \req{cs1}.
Equating the results of \req{cs1} and \req{cs2} allows one to solve
for the parameters $b_i$. The leading examples follow. 

\begin{center}
Table 1  The Asymptotic Expansion of $b$ as a function of $a$.
\end{center}
\bigskip
\begin{tabular}{ll} $m$ & relation between $a$ and $b$ \vtab \\
1 \tab & $b=2a-\frac{2}{3a}-\frac{19}{108a^3}-\frac{889}{9720a^5} -\dots$ \vhtab \\
2 & $b=8a-\frac{44}{15a}-\frac{2626}{3375a^3}-\frac{2157091}{5315625a^5} -\dots$
\vhtab \\
3 & $b=18a-\frac{657}{70a}-\frac{27080019}{10976000a^3}-\dots$ \vhtab \\
4 & $b=32a-\frac{1310}{63a}-\dots$ \vhtab \\
5 & $b=50a-\frac{8900}{231a}-\dots$ \vhtab \\
6 & $b=72a-\frac{1751309}{27456a}-\dots$ \\
\end{tabular}
\\[5pt]
The above establishes \req{asyj} to lowest order, and gives the first 
order corrections. Moreover, the definitions of $a$ and $b$ imply the relationship:
\beqn
b= \left(2GMl +  {2\over \gamma^2}(a-1)\right)^2.
\label{b vs a}
\eeqn

\begin{figure}[hbt]
\vspace{3cm}
\begin{center}
\leavevmode
\epsfxsize=12cm 
\epsfbox{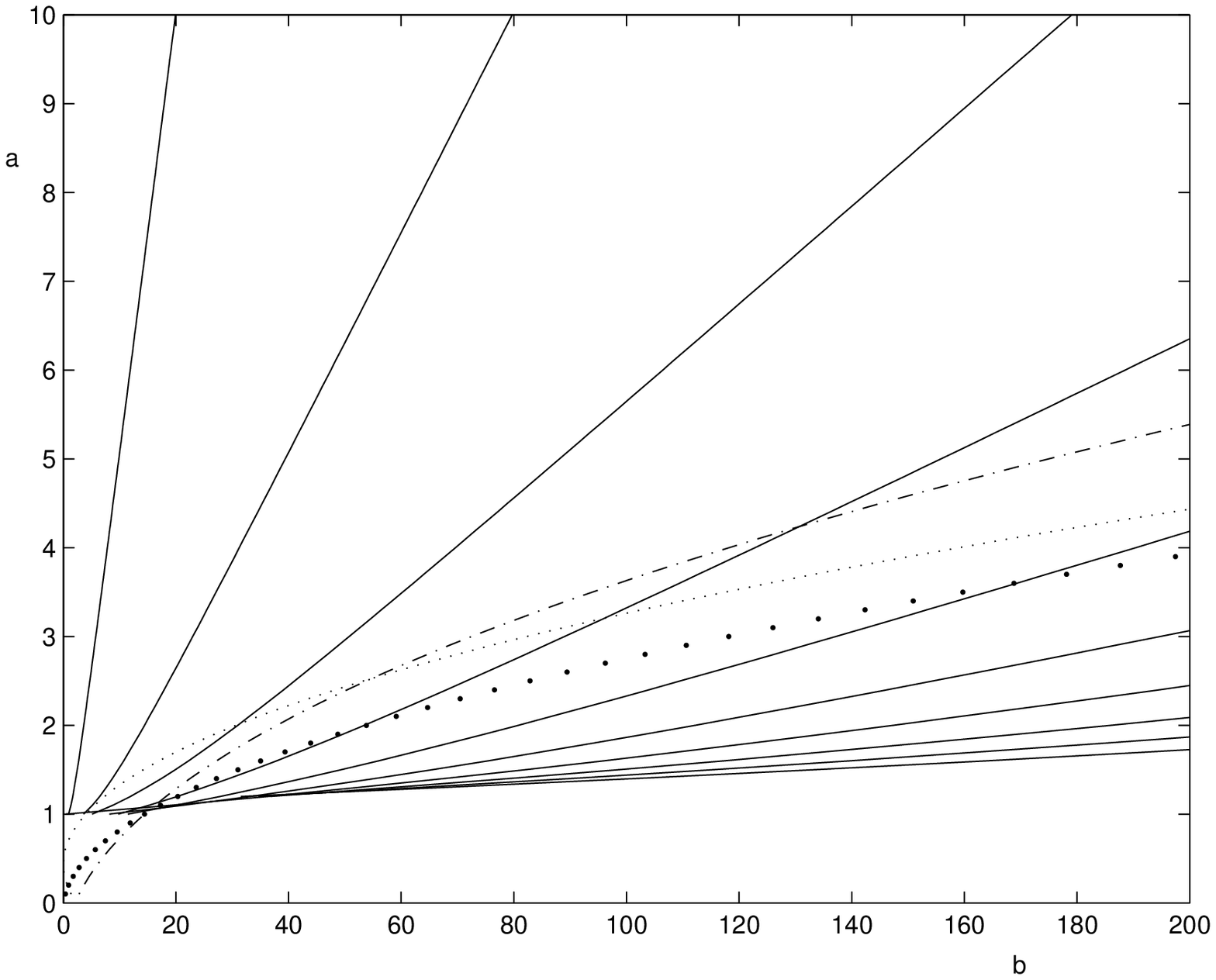}
\end{center}
\caption{Solid lines graph the relations between $a$ and $b$ from asymptotic expansions in Table 1. The parabolas graph (\ref{b vs a}) for various $M$ and $1/\gamma^2$. The large dotted line is $2GMl=1.90$, $1/\gamma^2=1.77$; the small 
dotted line is $2GMl=1.00$, $1/\gamma^2=1.77$; and finally the dot-dash line is $2GMl=1.90$, $1/\gamma^2=1.18$. The horizon is at $a=1$.  Intersection points 
of parabolas with the solid lines yield the spectrum for $j(\phi)$ either 
in terms of $a:=1+\frac
{1}{\gamma^2}(j(\phi)-2GM\ell)$ or of $b=(j(\phi))^2$.}
\label{Fig1}
\end{figure}

For fixed $M$ and $\gamma$,the expansions in Table 1 in conjunction with the relationship (\ref{b vs a}) can be used to determine the allowed spectrum for $j_i$, and consequentally $\phi(x_i)$, i.e. the discretization of the spatial slice. For large $\phi$ Table 1 confirms the approximate spectrum (\ref{asymptotic spectrum}), which approaches the continuum for large quantum numbers $m$. For small $\phi$ and hence $a$ close to 1, one needs to keep a suitable number of terms in the corresponding asymptotic expansions.

Figure 1 shows qualitatively the effects of changing the mass $M$ of the black hole and the discretization scale $\gamma$. Increasing the mass shifts the parabola down (small to large dots) and the spectrum for $a$ is squeezed closer to the horizon ($a=1$). Decreasing $\gamma$ (dot-dash to large dots) flattens the parabola and moves the eigenvalues of $b=(j(\phi))^2$ closer together. This behaviour corresponds to our physical expectation that the spatial topology near the horizon becomes more continuous with increasing mass or decreasing $\gamma$.

Note that the mass is expressed in terms of Planck units. For very large $M$, one expects the spectrum in the exterior of the black hole to be approximately continuous even near the horizon. This in fact can be seen to happen by noting from (\ref{b vs a}) that as $MGl$ increases, the horizon $a=1$ occurs for larger and larger $b$, which in turn implies that the parabola defined by (\ref{b vs a} first intersects the asymptotic expressions in Figure 1 for larger and larger $m$, where the lines are more and more dense. Thus the lowest eigenvalues get closer and closer together as $MGl$ increases.

\section{Conclusion: The Quantum State of a Black Hole}
We have presented a quantization scheme for the spacetime surrounding a generic, single horizon, spherically symmetric black hole. The key features of the scheme are:
\begin{itemize}
\item Only the spatial diffeomorphisms were fixed at the classical level; the choice of gauge fixing allows classical slices that are non-singular across the future horizon.
\item The single remaining constraint (the Hamiltonian constraint) does not contain any spatial derivates.  It describes the dynamics of a particle moving in a $1\over X^2$ potential, where the role of particle position is played by $X(x)=e^{\rho}$, the square root of the conformal mode of the metric.
\item The Hamiltonian constraint is quantized using Bohr quantization, which by construction yields a discrete spectrum for $X$ and induces a discrete spatial topology.
\end{itemize}

The interesting result is the induced discretization of the spatial topology exterior to the horizon. It is important to remember that our procedure specified the spectrum of the operator $\hat{X}$ at each spatial point. The Hamiltonian constraint then determined the spectrum for $j_i$, i.e. the discrete spatial topology.  Moreover, this topology has the desirable property that it approximates the continuum far from the horizon even for microscopic black holes, as well as near the horizon for macroscopic black holes.

We end on a speculative note. From Figure 1 it is clear that the parabolas intersect each line twice; the intersection furthest from $a=1$ determined the spectrum of $\phi$ to which we referred to above. However, the second intersection, closer to the horizon may also have physical significance. Note in particular that each parabola (i.e. for each value of $M$ and $\gamma$, there are an infinite number of intersections infinitesmally close to the horizon. This seems to imply that although the spatial topology just outside the horizon is discrete, there is in fact a continuum field theory residing on the horizon itself. At each spatial point in this (approximate) continuum theory, there is a set of $SO(2,1)$ generators that are defined classically in terms of the operators $X$ and $P$. It may be that the anomaly associated with the quantum version of this algebra near the horizon is connected to the statistical mechanical entropy of the black hole, much as in the work of \cite{carlip} and \cite{solodukhin}. 

There is of course a great deal yet to be done. The spectrum of $X$ and the resulting spatial topology below the horizon must be understood and the properties of the wave function at each spatial point should be analyzed. A complementary approach to the quantization of the same system, namely Schrodinger quantization is also being explored\cite{lk} and this should also yield insights into the physical interpretation. Most importantly, it is of great interest to see whether the present formalism can be implemented in the presence of matter, which would provide a new mechanism for analyzing the quantum dynamics of black hole formation and Hawking radiation. All this and more are currently under investigation\cite{quant_matter}.

\section{Acknowledgements} 

The authors are grateful to Ramin Daghigh, Arundhati Dasgupta, Viqar Husain, Jorma Louko and Vardarajan Suneeta for helpful conversations. This research is supported in part by the Natural Sciences and Engineering Research Council of Canada.

\end{document}